\documentclass[aps,prx,twocolumn,superscriptaddress,longbibliography]{revtex4-2}

\usepackage{amssymb}
\usepackage{amsmath}
\usepackage{graphicx}
\usepackage[dvipsnames]{xcolor}
\usepackage{bm}
\usepackage{dcolumn}
\usepackage{dsfont}
\usepackage[hidelinks]{hyperref}
\usepackage{physics}
\usepackage{siunitx}

\usepackage{cleveref}
\crefname{equation}{Eq.}{Eqs.}
\Crefname{equation}{Equation}{Equations}
\crefname{figure}{Fig.}{Figs.}
\Crefname{figure}{Figure}{Figures}
\crefname{section}{Sec.}{Secs.}
\Crefname{section}{Section}{Sections}
\crefname{appendix}{Appendix}{Apps.}
\Crefname{appendix}{Appendix}{Apps.}
\crefname{paragraph}{Sec.}{Secs.}
\crefname{table}{Table}{Tables}

\usepackage{multirow}
\usepackage{tabularx}

\usepackage{hhline}

\newcommand{\bb}{\hat b}
\newcommand{\bdag}{\hat b^\dagger}

\begin{document}

\title{Quantum optimal control of superconducting qubits based on machine-learning characterization}

\author{\'Elie Genois}
\email{elie.genois@usherbrooke.ca}
\address{Institut quantique \& D\'epartement de Physique, Universit\'e de Sherbrooke, Qu\'ebec J1K 2R1, Canada}
\author{Noah J. Stevenson}
\address{Quantum Nanoelectronics Laboratory, University of California, Berkeley, Berkeley CA 94720, USA}
\author{Noah Goss}
\address{Quantum Nanoelectronics Laboratory, University of California, Berkeley, Berkeley CA 94720, USA}
\address{Applied Math \& Computational Research Division, Lawrence Berkeley National Lab, Berkeley CA 94720, USA}
\author{Irfan Siddiqi}
\address{Quantum Nanoelectronics Laboratory, University of California, Berkeley, Berkeley CA 94720, USA}
\address{Applied Math \& Computational Research Division, Lawrence Berkeley National Lab, Berkeley CA 94720, USA}
\address{Materials Science Division, Lawrence Berkeley National Lab, Berkeley CA 94720, USA}
\address{Canadian Institute for Advanced Research, Toronto, Ontario M5G 1M1, Canada}
\author{Alexandre Blais}
\address{Institut quantique \& D\'epartement de Physique, Universit\'e de Sherbrooke, Qu\'ebec J1K 2R1, Canada}
\address{Canadian Institute for Advanced Research, Toronto, Ontario M5G 1M1, Canada}

\begin{abstract}
    Implementing fast and high-fidelity quantum operations using open-loop quantum optimal control relies on having an accurate model of the quantum dynamics. Any deviations between this model and the complete dynamics of the device, such as the presence of spurious modes or pulse distortions, can degrade the performance of optimal controls in practice. Here, we propose an experimentally simple approach to realize optimal quantum controls tailored to the device parameters and environment while specifically characterizing this quantum system. Concretely, we use physics-inspired machine learning to infer an accurate model of the dynamics from experimentally available data and then optimize our experimental controls on this trained model. We show the power and feasibility of this approach by optimizing arbitrary single-qubit operations on a superconducting transmon qubit, using detailed numerical simulations. We demonstrate that this framework produces an accurate description of the device dynamics under arbitrary controls, together with the precise pulses achieving arbitrary single-qubit gates with a high fidelity of $\sim$~\!99.99\,\%.
\end{abstract}

\date{\today}
\maketitle

\section{Introduction}\label{sec:Intro}
The precise characterization and control of quantum devices are central to the development of useful quantum technologies.
The powerful framework of quantum optimal control (QOC) theory can be used to go from a given description of the quantum dynamics, such as a characterized model, to realizing arbitrary quantum operations with maximal fidelity and minimal duration~\cite{peirce_optimal_1988,werschnik_quantum_2007,koch_quantum_2022}.
The successful practical implementation of many QOC approaches, such as GRAPE~\cite{khaneja_optimal_2005} and Krotov~\cite{krotov_global_1995}, thus rely on having an accurate model of the system dynamics to produce the desired output quantum state or process~\cite{boscain_introduction_2021}.
In simulation, optimizing the input controls using these methods can routinely yield quantum operations with decoherence-limited or even machine-precision fidelity, and these controls can have much shorter durations compared to what is achievable with simpler, monochromatic and flat, pulse shapes~\cite{motzoi_optimal_2011,egger_optimized_2013,rojan_arbitrary_2014,goerz_charting_2017,theis_simultaneous_2016}.

A critical problem with using open-loop QOC approaches in experiments is model bias, since the performance of the resulting controls is intrinsically limited by the underlying model accuracy.
Indeed, any mismatch between the model used to describe the quantum evolution and the actual dynamics of the physical system can cause the optimal controls to perform significantly worse in practice.
This performance degradation is routinely observed when considering parameter deviations to the assumed model, and has led to the development of robust QOC approaches~\cite{wesenberg_designing_2004,wang_composite_2012,buterakos_geometrical_2021,koswara_quantum_2021}.
These approaches sacrifice some control performance for the benefit of robustness under certain model parameter variations.
However, beyond being inaccurate with model parameters deviating from their ``true" values due to finite characterization precision and system drifts, the model used for QOC can additionally be incomplete, which also leads to important biases.
Out-of-model dynamics typically originate from unaccounted frequency- and power-dependent distortions in the control lines, crosstalk between qubits and control lines, and more generally coupling to spurious modes not included in the system modeling such as material defects, box modes, and neighboring couplers and readout apparatus.
Precisely characterizing and modeling the dynamics associated with each of these potential interactions is a challenging task, both experimentally and numerically. 
There is thus a need for alternative approaches to optimally controlling quantum systems.

The shortcomings of using inaccurate models in the control optimization has led to recent proposals of closed-loop optimization approaches based on reinforcement-learning (RL), which rely on direct interactions between a controller and the quantum system of interest~\cite{bukov_reinforcement_2018,borah_measurementbased_2021,sivak_modelfree_2022,porotti_deep_2022,porotti_gradientascent_2023}.
Using such a feedback control approach, the QOC problem can be made model-free, thus alleviating the aforementioned problem of model bias.
For example, this approach was applied experimentally to realize high-fidelity single- and two-qubit quantum gates, as well as a quantum error correction stabilization protocol in superconducting quantum devices~\cite{baum_experimental_2021,sivak_realtime_2023,ding_ftf_2023}.
Although useful for precisely calibrating a specific operation, these model-free approaches possess important drawbacks, notably in terms of sample efficiency, generalizability beyond performing a single quantum operation, ease of experimental implementation given the need for real-time feedback, and performance of the control solutions.
For instance, Porotti \textit{et al.}~\cite{porotti_gradientascent_2023} demonstrated that model-based gradient-ascent approaches, such as GRAPE, significantly outperform model-free RL approaches on standard state preparation optimal control tasks, both in terms of state fidelity and data efficiency.
This can be understood from the fact that in the model-free setting, the task of the learning agent is much more complex, as it needs to both resolve how a given control pulse (action) impacts the state or process fidelity (reward), and learn how to improve that control by navigating the enormous control parameter space.
The first part of this task can be recognized as a quantum characterization problem, whereas the second part of finding the optimal control strategy can be directly achieved using one of the many successful QOC approaches, instead of relying on the trial-and-error exploration of the RL approach.

Based on this understanding, we propose in this work to simplify the quantum optimal control problem by breaking it down into two parts that we perform in succession.
First, we frame the quantum characterization problem, or model learning problem, as a supervised machine learning (ML) task. We use a parametrized representation to learn a description of the quantum system dynamics directly from experimentally available data.
Second, we use that trained model in a gradient-based optimal control loop to find the external controls realizing our target operations.
In the following, we demonstrate that this modular and easily implementable approach can yield high-fidelity controls using experimentally realistic data, while providing notable benefits in terms of data efficiency, scalability to complex control problems, and device characterization.

The paper is organized as follows.
In~\cref{sec:Methods}, we describe our machine-learning-based optimal control approach and its benefits over alternative control approaches.
Using detailed numerical simulations, we then present a case study implementation of our approach and demonstrate its performance for realizing high-fidelity arbitrary single-qubit gates in a transmon qubit in~\cref{sec:SQ-gates}. 
In~\cref{sec:Usefulness}, we analyze the impact of out-of-model dynamics on the optimal control performance, and demonstrate the distinct advantages of our approach in the presence of model bias.
We conclude with a short discussion on the relevance of this work in~\cref{sec:Conclu}. 

\begin{figure}[t]
    \centering
    \includegraphics[width=\columnwidth]{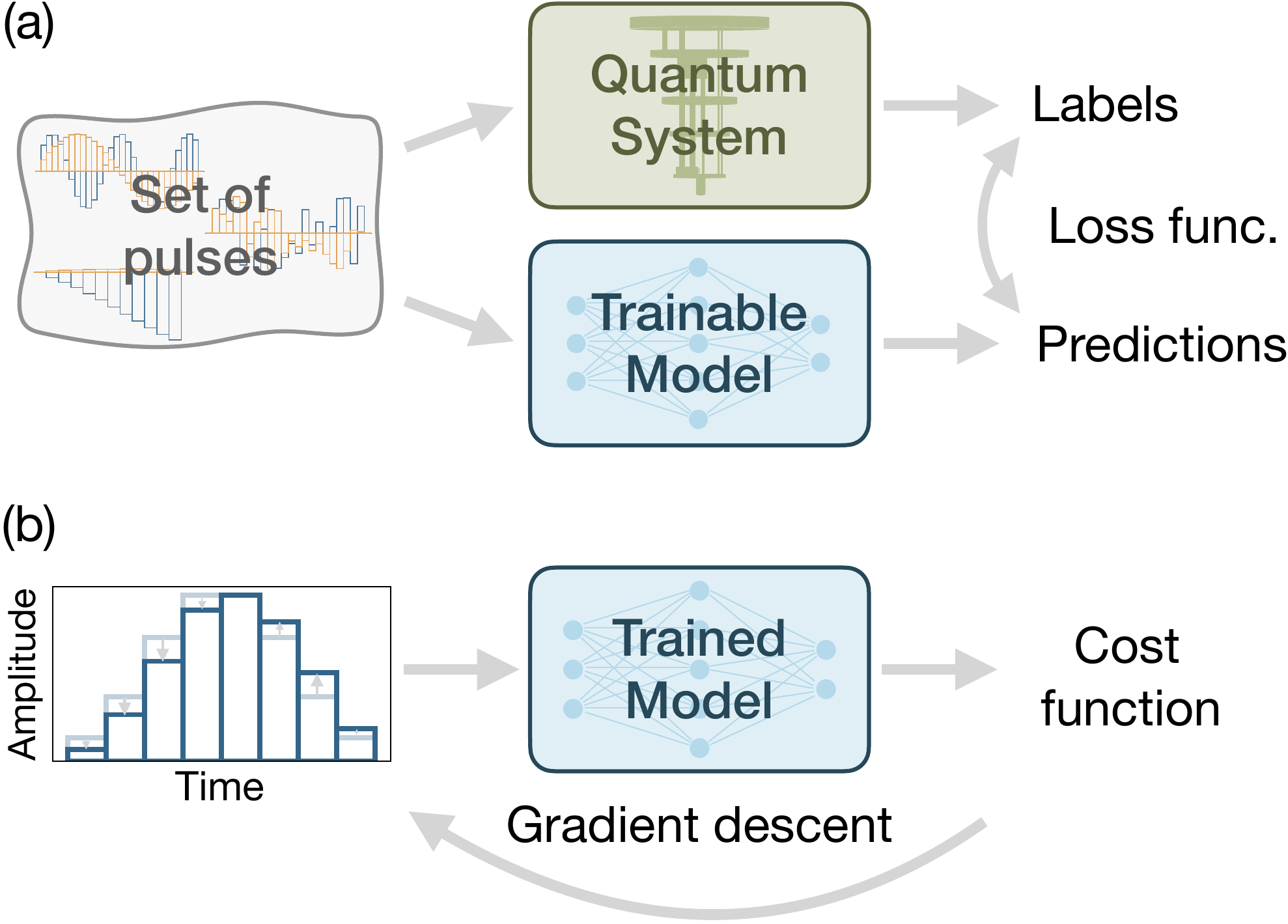}
    \vspace{-0.5cm}
    \caption{
        Optimally controlling a quantum system using machine-learning characterization.
        \textbf{(a)} A supervised machine learning approach is used to characterize the quantum dynamics produced by arbitrary input controls.
        The dataset is constructed by applying a set of diverse pulses sequentially to the quantum system and obtaining labels of the system observables using (projective) measurements.
        A parametrized representation (blue box) learns this transformation from input pulse shape to output quantum observables by minimizing a loss function related to the distance between its predictions and the data labels. 
        \textbf{(b)} The trained model is used directly in a gradient-based quantum optimal control loop to find the pulses best realizing the quantum operations of interest. The parametrized pulse shape is iteratively updated such as to minimize a cost function associated with a state or process infidelity.
    }
    \label{fig:Overview}
\end{figure}

\section{Machine-learning based QOC}\label{sec:Methods}
The two main steps of our proposed quantum optimal control approach based on machine-learning characterization are presented schematically in~\cref{fig:Overview}.
The characterization task consists of constructing a model that captures the transformation from arbitrary input pulse shapes to the resulting quantum state observables of interest. 
In this work, we focus on qubit gates and therefore take these observables to be the Pauli matrices on $n$ qubits, which are informationally complete to describe qubit states.
We can thus consider that the model we want to construct outputs the full quantum state $\rho(t)$. 
Note that a more restricted set of observables could be used if acquiring this information is prohibitive, for example in the case of a large Hilbert space, as long as the relevant metric to optimize can be described by the model output observables.

The idea of our approach is to learn this model directly from data taken on the device of interest using a supervised machine learning strategy.
To construct the dataset, we simply drive the quantum system sequentially with a large set of different pulses, and perform projective measurements to obtain bits of information about how the quantum state was transformed by each of these pulses.
After averaging over a chosen amount of shots, these measurement outcomes form the labels that the model will be trained to predict, using a loss function quantifying the distance between the model predictions and the labels. 
As detailed in~\cref{sec:SQ-gates}, additional physically motivated loss terms can be added to regularize the model's training, as was done in our previous work~\cite{genois_quantumtailored_2021}.
We emphasize that this machine-learning training is performed offline such that no real-time feedback between the model and the quantum system is required.

The second step consists of directly using the trained model in a gradient-based optimal control loop to find our control strategy.
At this point after the learning stage, the model parameters are fixed and only the parametrized pulse shapes are optimized.
Given the auto-differentiable nature of the model and its fast evaluation on graphics processing units (GPUs), these optimizations can be performed in less than a minute on standard hardware. Additionally, the optimization can be designed to include arbitrary experimental limitations in the cost function~\cite{leung_speedup_2017}, such as to directly find optimal controls that are easily implementable in practice.
Importantly, in contrast to direct feedback control approaches, the trained representation is not tied to a specific quantum gate and we can use the trained model to optimize for arbitrary quantum operations which can be described and realized by the same input controls and output observables as the model.
For instance, using a model outputting the quantum state of $n$ qubits, we can optimize for arbitrary qubit gates and qubit state preparations.
This can be directly generalized to higher dimensional systems such as qudits and bosonic modes, as long as these additional degrees of freedom are observable in the quantum system of interest.

An important advantage of our two-step approach to quantum control is that it is highly modular and customizable.
First, having a trained model describing the system dynamics for arbitrary input controls that is fast to evaluate can be very valuable for understanding, calibrating, and improving the quantum system of interest.
Indeed, this model can be used as a heuristic digital twin of the device to perform simulations and prototype new protocols. This is especially useful given that the trained model heuristically includes experimentally-relevant imperfections that were learned from data and that are typically not accounted for when modeling the system.
Additionally, this model is fully differentiable which means one can replace typically used parameter sweeps or gradient-free approaches with gradient-based optimization of the controls, which are faster and yield better solutions~\cite{nocedal_numerical_2006,machnes_gradient_2018}.

The second aspect making our approach modular is that the specific architecture used to learn from data, together with the optimal control algorithm used subsequently, can be explicitly engineered to satisfy the specific data, speed and performance requirements of a given quantum system.
This is particularly useful to explore the bias-variance tradeoffs of learning an accurate representation of the device dynamics and to perform model selection.
For instance, using a black-box learning model to remove any potential bias might require too much data to be trained to reach the desired precision, whereas a fully principled approach based on a physics description might be computationally prohibitive to model or unable to describe the dynamics accurately enough. 
Using a graybox approach~\cite{youssry_experimental_2024} combining physical priors about the device properties together with additional freedom might be optimal in such a realistic scenario, which can be directly realized within our framework.
Indeed, our approach allows us to use as much prior knowledge as desired in the learning model, anywhere from a parametrized master equation~\cite{krastanov_unboxing_2020,genois_quantumtailored_2021} to a fully general neural network, as long as the model outputs are informationally-complete for quantifying the desired quantum operations fidelity.

We note that approaches of directly combining characterization (system identification) and control have been studied in control theory~\cite{albertos_iterative_2002} and applied to quantum systems using a variety of models and data~\cite{wu_dGRAPE_2018,krastanov_stochastic_2019,chen_combining_2020,wittler_integrated_2021,ball_softwaretoolsQCTRL_2021,fyrillas_scalable_2024,youssry_experimental_2024}.
Focused on experimental feasibility and on mitigating the problem of model bias, this work provides a comprehensive framework for applying such an approach to superconducting and other solid-state devices and demonstrates its performance through detailed simulations. 
In addition, an important distinction of our work is the demonstration for the first time of using a trained neural network as the model for performing open-loop quantum optimal control, which allows us to significantly mitigate any model bias in practice.
In contrast to the similar approach by \textcite{youssry_experimental_2024}, we show that learning an explicit Hamiltonian description of the dynamics is not required for performing arbitrary QOC with high fidelity.

\section{Transmon single-qubit gates}\label{sec:SQ-gates}
We now present a comprehensive case study of our approach applied to realizing fast and high-fidelity microwave single-qubit gates in a fixed-frequency superconducting transmon qubit~\cite{koch_transmon_2007}.
We emphasize that our approach is agnostic to the specific physical implementation and could be applied to other platforms following the steps detailed here.
The available qubit controls in the chosen superconducting qubit system are microwave pulses generated at room temperature by classical control electronics, which travel down a cryogenic fridge before reaching the qubit.
Information about the dynamics resulting from the controls can be acquired via projective measurement of arbitrary qubit operators using standard dispersive qubit readout~\cite{blais_cavity_2004, blais_RMP_2020}.
Given these input controls and available output observables, the task of interest is to design the external microwave pulses such as to realize arbitrary qubit unitaries with the highest fidelity and minimal duration.

The following four subsections detail the four steps required for achieving this task using our machine-learning based quantum optimal control approach (MLQOC). Namely, the framework consists of (1)~acquiring a dataset on the device, (2)~training a machine-learning model on this data, (3)~performing QOC optimizations using this trained model, and (4)~testing the performance of the optimal controls on the device of interest.
As a proof of principle of the method, here the data set is generated by numerical simulation of the device (step 1), and the same is true for testing the performance of the optimized controls (step 4).
The other two steps remain unchanged when working with an experimental device.
In our simulations, we take special care not to make approximations which would artificially simplify the model learning and optimal control tasks considered, in an effort to provide a convincing demonstration of our approach in an experimentally realistic scenario.

\subsection{Simulation dataset}\label{subsec:simulation}

\textit{Physical model.}
We consider a transmon qubit capacitively coupled to a microwave drive line described by the Hamiltonian ($\hbar=1$)~\cite{koch_transmon_2007}
\begin{align}
    \hat H_\mathrm{tot}(t) &= 4E_C(\hat{n}- n_g)^2 - E_J\cos{\hat{\varphi}} + \Omega(t)\hat{n}\\
    &= \hat H_\mathrm{trans} + \Omega(t)\hat{n}\,,
    \label{eq:hamiltonian}
\end{align}
with $E_C$ ($E_J$) the charging (Josephson) energy, $\Omega(t)$ the applied microwave drive coupling to the transmon charge operator $\hat{n}$, and $\hat{\varphi}$ the canonically-conjugate phase operator.
Given that the qubit is operated in the transmon regime, $E_J/E_C=110$, and that we are interested in quantum operations limited to the qubit subspace, we can safely ignore the gate charge $n_g$~\cite{koch_transmon_2007}.
Throughout this work, we simulate this full Hamiltonian keeping the first 5 eigenstates of $\hat H_\mathrm{trans}$.
We do not perform any rotating-wave approximation (RWA) in order to capture the fast-oscillating dynamics that are relevant for high-fidelity and fast operations~\cite{motzoi_optimal_2011,shillito_fast_2021}.

We parametrize the drive pulses $\Omega(t)$ as the input to the arbitrary waveform generator (AWG) that produces in-phase ($S_I$) and quadrature ($S_Q$) signals.
These signals are then combined to a local oscillator (LO) of frequency $\omega_\mathrm{LO}$ to be up-converted to the desired GHz-frequency signals, a process known as sideband mixing~\cite{krantz_quantum_2019,baur_realizing_2012}. 
This pulse parametrization reflects the actual controls of a realistic experiment, allowing our model to capture potential IQ-mixer imperfections and other pulse distortions, in addition to allowing for precise control of the pulses frequency spectrum.
In the absence of imperfections in the control electronics and pulse distortion in the control lines, the microwave pulse reaching the qubit is described by
\begin{align}
    \Omega(t) = S_I(t) \cos(\omega_\mathrm{LO}t) + S_Q(t) \sin(\omega_\mathrm{LO}t).
    \label{eq:awg-signals}
\end{align}
To drive the transmon on resonance at $\omega_\mathrm{q}$ and avoid IQ-mixer imperfections producing distorted output signals at frequencies close to $\omega_\mathrm{LO}$, the AWG signals are typically generated by convolving the pulse envelope with an intermediate frequency oscillation at frequency $\omega_\mathrm{IF}$ such that $\omega_\mathrm{LO}+\omega_\mathrm{IF}=\omega_\mathrm{q}$~\cite{baur_realizing_2012}. 
We reproduce this experimental condition here using $\omega_\mathrm{IF}/2\pi=$~\SI{100}{\mega\hertz} and $\omega_\mathrm{LO}/2\pi=$~\SI{6.198}{\giga\hertz} to reach the qubit frequency at $\omega_\mathrm{q}/2\pi\approx$~\SI{6.298}{\giga\hertz}.

As shown in~\cref{appendix:IQmixer}, when restricting the description to the computational states of the transmon, the effect of the drive $\Omega(t)$ takes the usual form in the rotating frame of the qubit, $\hat H_\mathrm{qubit} = I(t) \hat\sigma_x + Q(t) \hat\sigma_y$, for a resonant drive and under the rotating-wave 
approximation. The real-valued signals $I$ and $Q$ can be expressed as linear combinations of the AWG signals $S_I$ and $S_Q$.
We thus understand how controlling $S_I$ and $S_Q$ allows us to perform arbitrary single-qubit gates.
Here, we avoid the approximations mentioned above and simulate the full transmon Hamiltonian~\cref{eq:hamiltonian} directly using~\cref{eq:awg-signals} with arbitrary time-dependent signals $S_I$ and $S_Q$.

In the AWG as in our simulations, the pulse shapes $S_I$ and $S_Q$ are specified by real amplitudes positioned at a finite number of times during the operation, called \textit{pixels}.
We take the size of these pixels to be $\SI{1}{\nano\second}$.
Importantly, we apply a Gaussian filter to these discrete signals to (i)~interpolate between pixels and simulate experimentally-accurate continuous time evolutions beyond a piece-wise constant pulse approximation~\cite{motzoi_optimal_2011,machnes_gradient_2018}, and (ii)~account for the finite bandwidth of the AWG and the filters used in experiments. We use a Gaussian filter standard deviation of $\SI{250}{\mega\hertz}$.

To simulate the full time dynamics of the system under the application of the drives, we solve the Lindblad master equation (ME) 
\begin{align}
    \frac{\mathrm{d}\hat{\rho}}{\mathrm{d}t} = -i \left[\hat H_\mathrm{tot}(t), \hat{\rho}\right] 
    + \gamma\mathcal{D}[\bb]\hat{\rho}
    + 2\gamma_\varphi\mathcal{D}[\bb^\dag\bb]\hat{\rho},
\end{align}
which accounts for transmon relaxation and dephasing~\cite{blais_RMP_2020}. In this expression, $\gamma$ ($\gamma_\varphi$) is the relaxation (pure dephasing) rate, and $\mathcal{D}[\hat{X}]\hat{\rho} = \hat{X}\hat{\rho}\hat{X}^\dag - \{\hat{X}^\dag\hat{X},\hat{\rho}\}/2$ the Lindblad dissipator.
In the following, we use relatively high relaxation and coherence times of $T_1=T_2=$~\SI{300}{\micro\second}, corresponding to $\gamma/2\pi=$~\SI{3.33}{\kilo\hertz} and $\gamma_\varphi/2\pi=$~\SI{1.67}{\kilo\hertz}.
This choice allows us to resolve the finite control precision of our MLQOC approach, beyond the gate fidelity limit set by decoherence.
We use the open-source library \texttt{dynamiqs} to perform these simulations, which alows us to use GPU acceleration and to efficiently batch the simulations over multiple pulse shapes in parallel~\cite{guilmin_dynamiqs_2024}.

\begin{figure*}[t]
    \centering
    \includegraphics[width=\textwidth]{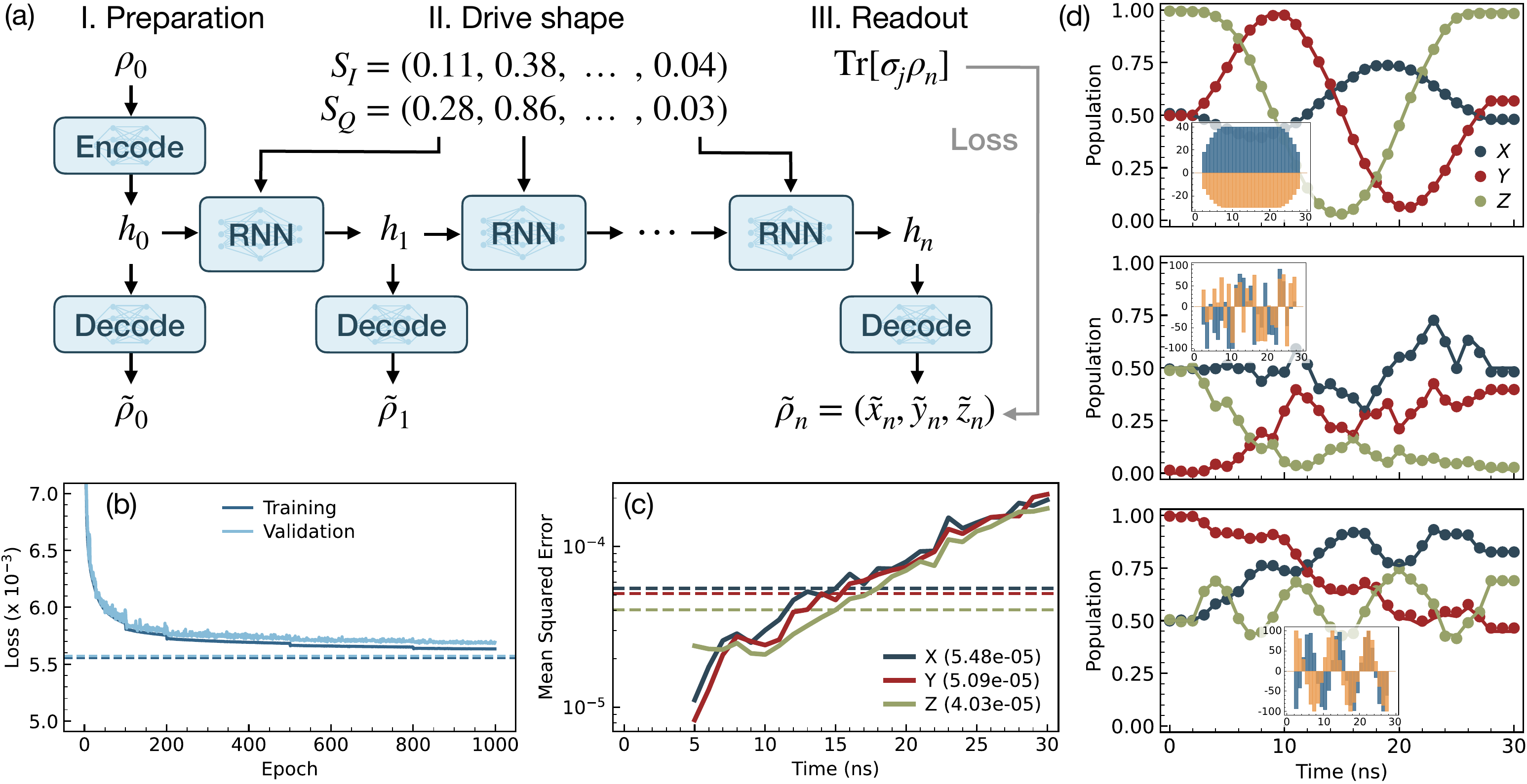}
    \vspace{-0.5cm}
    \caption{
        Learning driven qubit dynamics using supervised machine learning.
        \textbf{(a)} Model architecture. Preserving the structure of the three-step experiment, the model uses a small fully connected neural network (FNN, \textit{Encode}) to map the nominally prepared quantum state ($\rho_0$) to the initial hidden state representation of the model ($h_0$). Each pixel of the input pulse shape is fed sequentially to a recurrent neural network cell (\textit{RNN}). A second FNN (\textit{Decode}) is used to produce the predicted quantum state $\tilde\rho_n$, here parametrized using Pauli expectation values. During training, the experimental readout outcomes are used as labels and the model parameters are adjusted by minimizing a loss function, here principally composed of a mean squared error (MSE) between the labels and predictions.
        \textbf{(b)} Loss on the training and validation datasets during a supervised LSTM training. All hyperparameters and dataset characteristics are detailed in~\cref{appendix:MLtrainings}. The dashed lines correspond to the sampling noise floor of the noisy labels for both datasets, see text for details.
        \textbf{(c)} Mean squared error of the predicted quantum state expectation values as a function of pulse duration, averaged over the entire test dataset. Dashed lines correspond to the median over time, with values reported in the legend. The MSE is zero for the first 4~\SI{}{\nano\second} because we impose a zero padding of 2~\SI{}{\nano\second} at the beginning and end of every pulse.
        \textbf{(d)} Sample model predictions (circles) and true quantum trajectories (full lines) for three input pulse shapes unseen during training. The envelopes of the gaussian flat top (top), random (middle) and sinusoidal (bottom) pulses are shown in inset, with axes corresponding to drive amplitude in~\SI{}{\mega\hertz} and time in~\SI{}{\nano\second}. Note that during training, the ML model never has access to the true quantum trajectories used here for evaluating model performance in panels (c-d), but only to the noisy labels obtained from $N_\mathrm{shots}=32$ projective measurements.
    }
    \label{fig:Machine-learning}
\end{figure*}

\textit{Experiment and dataset generation.}
As illustrated in~\cref{fig:Machine-learning}(a), the experiment we consider consists of three steps, where (i) the transmon is prepared in one of the six cardinal states of the Bloch sphere, (ii) a microwave drive with a given pulse shape and duration is applied to the qubit, and (iii) the qubit is readout in the basis $\sigma_j\in\{X,Y,Z\}$.
This same experiment is repeated $N_\mathrm{shots}$ times to acquire statistics, where $N_\mathrm{shots}\geq1$, and the label associated with this specific pulse shape and preparation is the qubit expectation value estimate resulting from averaging these shots. 
We emphasize that we never feed the full quantum state to the model and that this label, i.e.~the average measurement result, is a realistic noisy estimate of the true qubit expectation value that the model is trying to predict. 
State preparation and measurement (SPAM) errors can add noise to these labels. This noise can be made effectively unbiased using standard error mitigation strategies based on inverting the confusion matrix or on measuring half of the shots with the negative measurement operator. For simplicity in evaluating model performances, we did not model the effect of the mitigated SPAM errors which would typically be significantly smaller than shot noise in our simulations ($N_\mathrm{shots}=32$).

To construct the supervised ML datasets, this experiment is repeated for all of the pulse shapes in a chosen set of pulses, while randomizing over the prepared state and measurement axis for each of these pulses.
Implementation details for constructing the pulse set such as to efficiently sample the control space is presented in~\cref{appendix:datasets}.
We have found that using a combination of random pulse shapes together with physically motivated envelopes, such as Gaussian, sinusoidal, and flat-top envelopes with DRAG components~\cite{motzoi_simple_2009} was sufficient for the ML models considered to learn an accurate and generalizable representation of the quantum dynamics.

Putting this data together in the form of supervised learning datasets, each input consists of a one-hot encoded vector $\vec{p}$ describing which of the six cardinal states was prepared, together with a two-dimensional real array $\mathbf{S}=(\vec S_I, \vec S_Q)$ containing the pulse shape amplitudes at each pixel of the evolution for the two drive quadratures.
The output labels are a combination of a one-hot encoded vector $\vec{m}$ capturing which Pauli operator was measured together with a single floating point number representing the associated qubit expectation value estimate $\operatorname{Tr}[\sigma_j\rho(t)]$.
Finally, we split this dataset into training, validation and test sets, as is standard in supervised learning approaches to both avoid over-fitting by selecting the best model from the performance on the validation set, and to obtain unbiased model performance metrics by evaluating the chosen models on a separate test set~\cite{goodfellow_deeplearning_2016}.

\subsection{Model learning}\label{subsec:model-learning}
\textit{Model architecture.}
The parametrized representation we use to learn the quantum dynamics from the data described above is the physics-inspired neural-network illustrated in~\cref{fig:Machine-learning}(a).
It is principally composed of a recurrent neural network (RNN) architecture~\cite{goodfellow_deeplearning_2016,lipton_reviewrnn_2015}, which processes the input pulse shapes one pixel at a time such as to preserve the time-ordered structure of the learning problem.
The RNN, composed here of a long short-term memory (LSTM) unit cell~\cite{hochreiter_lstm_1997}, is performing the same set of operations at each time step in a recurrent fashion, using the new input together with a hidden state ($h$) to encode information about the context of that input. This vector allows the model to capture correlations within the input data and to potentially keep a memory of previous inputs.
Given our quantum characterization problem where the model needs to output the quantum state at different times, we engineer a direct correspondence between the hidden state of the model $h_n$ and the quantum state of the system $\rho_n$.
We thus use an encoding layer, a small fully-connected neural network, to map the initially prepared quantum state to the initial hidden state of the model, and use a similar decoding layer to map this hidden state back to the qubit state prediction $\Tilde{\rho}(t)$.

To have a model that can efficiently train on a finite and realistic dataset, we have designed the model architecture to explicitly preserve most of the structure of the physical problem at hand. Nonetheless, the model is general enough to go significantly beyond the usual assumptions of a Markovian, single-mode description of the qubit dynamics.
As demonstrated in~\cref{sec:Usefulness}, this choice is motivated by our objective of obtaining high-fidelity controls in scenarios where the physical description used in typical open-loop control approaches is insufficient.
We emphasize that many other architecture choices could be directly used within our framework, such as a transformer model~\cite{vaswani_attention_2017} or a parametrized master equation.

To train the model, we use a loss function principally composed of a Mean Squared Error (MSE) loss between the model predictions and the labels, as is common for regression tasks in supervised machine learning.
Additionally, following Ref.~\cite{genois_quantumtailored_2021}, we expand the loss function with terms assuring that the RNN outputs are valid quantum states, i.e.~that they are positive, and that the initial state predicted by the model corresponds to the known prepared state of the qubit. These physically motivated loss terms help regularize the model training, as it explores the parameter space of valid representations more efficiently~\cite{raissi_pinn_2019}.
An explicit expression for the full loss function, together with the hyperparameters used, can be found in~\cref{appendix:MLtrainings}.

\textit{Quantum characterization results.}
A typical training of the RNN model is presented in~\cref{fig:Machine-learning}(b) on a training dataset comprising of 3.2 million pulse shapes of maximal duration of 30~ns, each measured for 32 shots.
As demonstrated in~\cref{appendix:data-requirements}, similar performances can be reached using significantly less data.
Both the training and validation MSE losses reach a value close to the sampling noise floor of about $5.6\times10^{-3}$, which is obtained by computing the same MSE metric assuming perfect knowledge of the qubit expectation values, as given by the simulation data.
This imprecision is significant as it corresponds to an average distance between the noisy labels and the model output qubit state probabilities of about 7\,\%. Such a high noise floor raises the question of whether the ML model is learning an accurate description of the quantum dynamics.

\begin{figure*}[t]
    \centering
    \includegraphics[width=\textwidth]{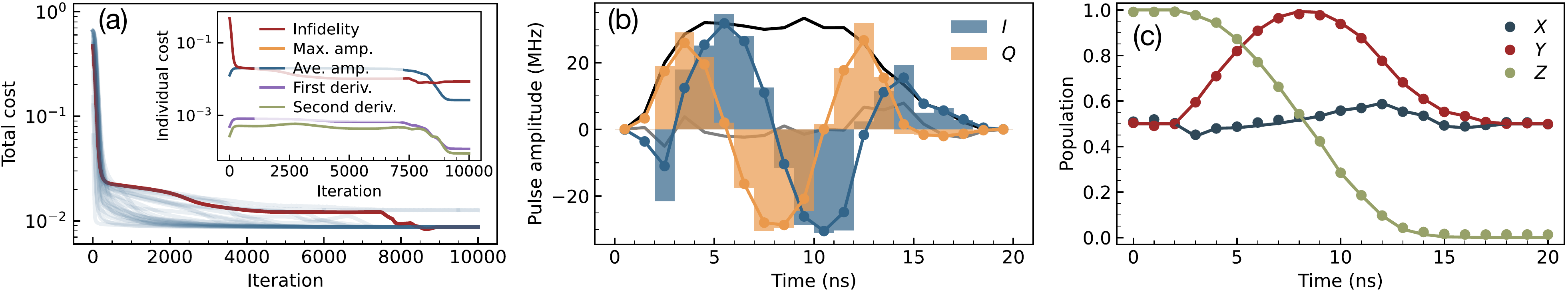}
    \caption{
        Quantum optimal control of a \SI{20}{\nano\second} $R_X(\pi)$ gate on a trained RNN.
        The optimal MLQOC pulse fidelity is 99.994\,\% on the 5-level transmon.
        \textbf{(a)} Parallel optimization of 30 randomly initialized pulse shapes. The optimization of the best performing pulse is highlighted in red, with the inset showing its decomposition into individual costs.
        \textbf{(b)} Resulting raw (bars) and gaussian filtered (dots with line) optimal pulse shape for the two drive quadratures $I$ and $Q$.
        \textbf{(c)} Predicted (dots) and true (lines) quantum state evolution when applying the optimal pulse shape on the initial state $|1\rangle$. 
    }
    \label{fig:QOC}
\end{figure*}

Leveraging the fact we are working with simulation data, we can go beyond the experimentally-available noisy labels and compare the model predictions directly to the ground truth quantum trajectories obtained from simulation.
In~\cref{fig:Machine-learning}(d), we can qualitatively observe the high accuracy of the trained model at describing the qubit dynamics for various input pulse envelopes, as the model predictions (dots) match closely the true trajectories (full lines).
This agreement is quantified in~\cref{fig:Machine-learning}(c), showing a median MSE of about $5\times10^{-5}$ across the three measurement axes and over different pulse lengths on the previously unseen test dataset.
This order of magnitude improved accuracy on the predicted outcome probabilities indicate that, remarkably, the ML model is able to use the noisy labels that one can obtain experimentally in order to learn the mapping from arbitrary pulse shapes to highly accurate quantum dynamics.
The error associated with the model predictions is lowest at the initial time where we have the most accurate information about the qubit state given the finite number of prepared states shared amongst all experiments.
The finite model precision for evolving the quantum state for every pixel then leads to a linear increase in the prediction error over time, which is translated into a quadratic increase on the MSE in~\cref{fig:Machine-learning}(c).
This behavior is expected for any model with a finite accuracy, and is akin to evolving the state using the ME with slightly inaccurate model parameters, or to numerically solving a differential equation with a finite precision solver.

The accuracy of the ML model can be significantly improved by extending the training dataset, for example using more shots such as to construct more precise labels, and performing more experiments with different pulse shapes such as to explore more of the control space. 
As we will show next, the amount and quality of the training data used here is sufficient to obtain high-fidelity controls from the model.
Additional results exploring how the quality and quantity of training data impact the learned representation and the performance of the resulting optimal controls are presented in~\cref{appendix:data-requirements}.

The neural network model takes about an hour to train on a standard Nvidia RTX 3080 GPU.
This time could be significantly reduced if needed for a specific implementation, for example by pre-training the model on simulation data before training on the experimental measurements.
This timescale is comparable to the experimental data acquisition time for building the datasets using transmon qubits, which is a few hours per million pulse shapes with 32 shots, including the overhead of pulse sequence uploading on the control electronics.

\subsection{Quantum Optimal Control}\label{subsec:qoc}
Having demonstrated that a RNN can be used to accurately learn quantum dynamics from experimentally realistic data, we now present how one can successfully use that representation to perform quantum optimal control.
We focus on single-qubit quantum gates and optimize the external controls such as to maximize the average gate fidelity of a given unitary $U_\mathrm{target}$.
As schematically illustrated in~\cref{fig:Overview}(b), the optimization procedure simply consists of making each pixel of the input pulse a free parameter and computing the evolution of the quantum system due to these controls using the trained ML model with fixed internal parameters.
By evolving a set of initial quantum states forming a unitary 2-design, one can compute the average gate fidelity of the arbitrary black-box quantum channel~\cite{dankert_2design_2009,debroy_cafe_2023}, which is represented here by a neural network.
For the case of one qubit, the six cardinal states used as preparations in our experiment form a unitary 2-design~\footnote{More generally, a set of $4^n$ states are required to form a unitary 2-design on $n$ qubits.}.
Finally, an optimization algorithm is used to update the input controls such as to maximize the fidelity of the operation of interest.

Making use of the fact that the RNN is fully auto-differentiable, we employ a gradient-based approach and design the cost function to include a multitude of experimentally-relevant constraints, without having to analytically derive an expression for their gradient.
In addition to the cost associated with the average gate infidelity of the operation of interest, we use regularizing costs that yield smooth and low amplitude controls, which are desirable in the experiment to mitigate crosstalk and pulse distortion effects~\cite{leung_speedup_2017,heeres_implementing_2017}. 
In particular, we use a cost term proportional to the average absolute pulse amplitude to penalize for unnecessarily strong pulses, a cost for amplitudes $|\Omega|/2\pi>$~\SI{100}{\mega\hertz}, together with terms proportional to the first and second derivatives of the pulse shapes to find smooth controls with a limited frequency spectrum.
The full cost function and optimization parameters used here are presented in~\cref{appendix:QOC}. 

A typical control optimization on the trained RNN is shown in~\cref{fig:QOC}(a) for the case of a \SI{20}{\nano\second} $R_X(\pi)$ transmon qubit gate.
Benefiting from the fast forward and backward evaluation of the RNN, 30 randomly initialized pulses are optimized in parallel on a single GPU card in under a minute.
This batched optimization and ability to perform a large number of iterations significantly mitigates the convergence issues that many GRAPE-like approaches face.
As a result, we can use the trained model to obtain optimal controls for a wide variety of gates on a very short timescale, which can be used to create a universal gate set of optimal pulses or to compile continuously parametrized gates.

The optimal control found by our MLQOC approach yields a 99.994\,\% average gate fidelity on the full 5-level transmon model.
The pulse shape and resulting qubit dynamics are illustrated in~\cref{fig:QOC}(b-c).
Importantly, this optimal pulse is smooth and easily implementable in practice with limited bandwidth electronics.
The main sinusoidal oscillations seen in the optimal pulse correspond to the expected intermediate frequency of about 2$\pi\times$\SI{100}{\mega\hertz}.
Deconvolving this intermediate frequency, we obtain a smooth envelope in the quadrature effectively driving $\sigma_X$ (black line), together with a small orthogonal component (gray line) necessary to adjust the $Z$ phase of the resulting unitary and to mitigate leakage to the $|2\rangle$ state~\cite{motzoi_simple_2009}.
As shown in panel (c), the pulse produces coherent dynamics on the qubit that resemble the one of standard sinusoidal and Gaussian control pulses, both according to the RNN model (dots) and the true master equation dynamics (lines). 

\subsection{MLQOC performance}\label{subsec:mlqoc}
An important advantage of our approach is that the trained model is not tied to a specific set of gates and can be used to optimize for arbitrary quantum operations captured by the model inputs and outputs, here any single-qubit unitary.
As a demonstration, we use the same trained RNN model presented in~\cref{fig:Machine-learning} to optimize for a variety of gates and compile the results in~\cref{tab:QOC}.
We first optimize for $\pi$ and $\pm\pi/2$ rotations around the three axes of the Bloch sphere.
This gate set, together with the identity operation $I$ realized by applying no drive, generates the full single-qubit Clifford group. 
We obtain an average fidelity of about 99.99\,\% for this gate set, demonstrating that finite-precision model trained on realistic data can yield high-fidelity quantum operations.
We have also optimized 100 unitaries sampled uniformly at random from the Haar measure~\cite{mezzadri_haar_2007}, and obtained gates with similar performances.
This result further demonstrates that the trained RNN representation is accurately capturing arbitrary dynamics on the qubit subspace of our transmon model such that it can be used directly for performing QOC.

The fact that the trained ML model does not fully reach the coherence-limited average gate fidelities of $\sim~\!99.997$\,\% can be attributed to the finite precision of the heuristic that the RNN learns to represent the quantum dynamics.
As shown in~\cref{appendix:data-requirements}, reducing shot noise and increasing the training dataset size does not significantly improve the control fidelity results, and important stochasticity in the performance of similar models remains, even for trained models performing the quantum characterization task significantly better than the one presented in~\cref{fig:Machine-learning}.
Given that it is possible to train a model that would yield coherence-limited gates, such as a model approaching the master equation used to generate the data, it would be interesting to try refining the training data by sampling around the optimal pulses or to explore the performance of different machine-learning architectures.
Despite this limitation, we show in the next section that our MLQOC approach can significantly outperform open-loop QOC approaches in practical scenarios by alleviating the problem of model bias.

\begin{table}[t]
  \def\arraystretch{1.4}
  \begin{tabularx}{1\columnwidth}{>{\centering\arraybackslash}X | >{\centering\arraybackslash}X | >{\centering\arraybackslash}X}
    \textbf{SQ gate} & \textbf{Duration (ns)} & \textbf{Fidelity (\%)} \\ \hline \\[-1.5em]
    $R_{X,Y,Z}(\pi)$ & 20 & 99.993 \\
    $R_{X,Y,Z}(\pm\pi/2)$ & 15 & 99.983 \\
    Haar random & 20 & 99.984 \\[0.2em]
  \end{tabularx}
  \caption{
    Gate performance of the MLQOC approach, evaluated on the full 5-level transmon. The trained model can be used to find arbitrary Clifford and Haar random unitaries with high fidelity. 
    Reported fidelities are the average over the different axes ($X$, $Y$, $Z$) for Clifford gates and the median over 100 uniformly sampled gates for the Haar random gates.
    The transmon coherence limits are 99.9975\,\% and 99.9967\,\% for \SI{15}{\nano\second} and \SI{20}{\nano\second} gates, respectively.
    Note that the gate durations include a \SI{4}{\nano\second} zero padding to avoid gate bleed-through.
    }
  \label{tab:QOC}
\end{table}

\begin{figure}[t]
    \centering
    \includegraphics[width=\columnwidth]{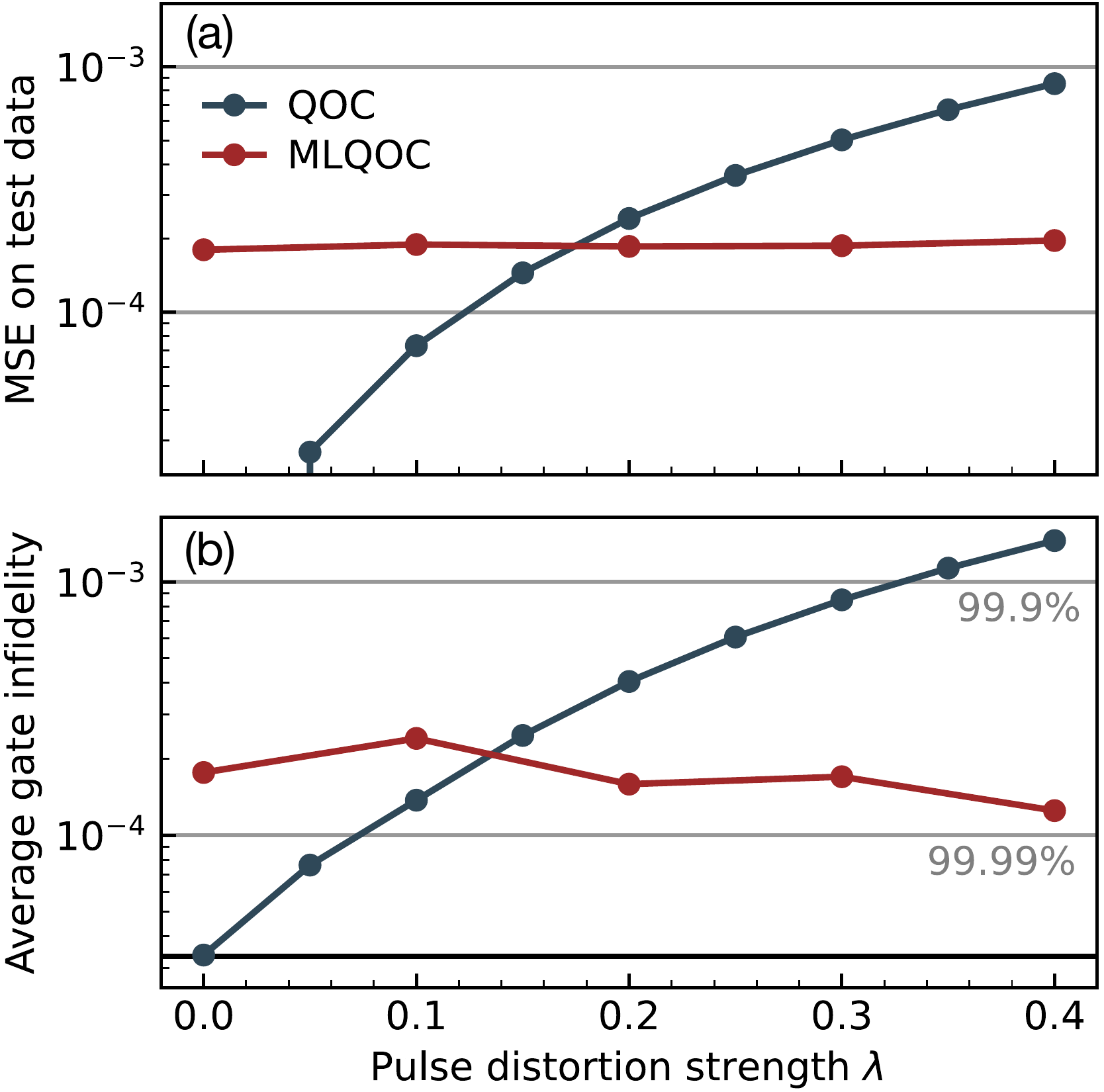}
    \caption{
        Demonstrating the robustness of the proposed MLQOC approach under a realistic source of model bias in the form of random pulse distortions.
        \textbf{(a)} Predicting the quantum state dynamics of the test dataset using the master equation model (blue) and RNN models trained on experimentally-available data containing 1 million pulse shapes, each sampled 32 shots (red).
        \textbf{(b)} Performance of the \SI{20}{\nano\second} $R_X(\pi)$ gates found by optimizing on the models of panel (a), as evaluated on the full transmon model.
        The gate coherence limit is indicated by a black line.
    }
    \label{fig:Robustness}
\end{figure}

\section{Comparison with open-loop QOC}\label{sec:Usefulness}
Having presented proof-of-principle results of our approach on transmon single-qubit gates, we now demonstrate the advantage of using MLQOC in realistic experimental settings where model bias will necessarily be present.
In particular, we show that the MLQOC approach studied here achieves significantly better quantum gates than typical open-loop QOC methods under realistic model bias.
This improvement is achieved by learning an accurate representation of the device dynamics from data, and using that model to design high-fidelity controls.

We consider model bias solely coming from out-of-model dynamics, i.e.~effects that are present in the quantum device but not in the physical modeling used by the open-loop QOC approach.
We take these dynamics to be unaccounted pulse distortions in the control lines, which produce discrepancies between the pulse shape the qubit receives and the one we model.
Note that considering pulse distortions is simply a choice for demonstrating the performance of MLQOC in the presence of model bias, although this choice is motivated by realistic experimental scenarios with superconducting qubits~\cite{gustavsson_improving_2013,rol_cryoscope_2020,lazar_calibration_2023,hyyppa_reducing_2024}.
The methodology used to generate the random pulse distortions is detailed in~\cref{appendix:distortions}.

In~\cref{fig:Robustness}, we present the performance of the open-loop (blue) and machine-learning-based (red) QOC approaches at performing the two tasks identified within our framework, namely the quantum characterization as reported by the MSE in panel (a) and the optimal control as reported by the obtained average gate infidelity in panel (b), both as a function of the amplitude of the random pulse distortions $\lambda$. Scaling this amplitude $\lambda$ directly translates into scaling the magnitude of the model bias.
For instance, a distortion of $\lambda=0.2$ corresponds to pulses deviating on average by only 3\,\% from their original shape, which is less than \SI{0.3}{\mega\hertz}.

Since the model used for the open-loop QOC approach is the full master equation used to describe the transmon qubit, it perfectly describes the quantum system when no model bias is present ($\lambda=0$), and achieves coherence-limited single-qubit gates.
However, as we scale the amplitude of the pulse distortions ($\lambda>0$), the mean-squared error for predicting the true quantum states of our test dataset grows sharply. Consequently, the controls optimized using the ME model yields single-qubit gate fidelities that rapidly drop below 99.9\,\% when amplifying the out-of-model dynamics, see panel (b).

In contrast, our proposed ML framework is designed to learn the entire transformation describing our control of the quantum system, from the input pulses of the AWG all the way to the measurement bits we extract, which includes pulse distortions as well as any other dynamics of the quantum device that can affect our measurements.
As shown in panel (a), the ML model precision is then virtually unaffected by the added pulse distortions, simply because these dynamics are present in the data it learns from.
Using these trained models in panel (b) then allows us to find optimal controls that remain close to 99.99\,\% fidelity, even under important pulse distortions.

These results constitute a clear demonstration that in a realistic scenario where model bias limits our ability to perform open-loop QOC, our approach of employing ML can lead to a considerable gain in quantum control fidelity.
We note that even in a well characterized system where our physical model yields similar quantum characterization performance to the trained ML model, our MLQOC approach can still offer an advantage. This is the case here for $\lambda=0.2$ where the MSE are similar for both approaches, see panel (a), but the resulting fidelity is superior for MLQOC, see panel (b).
This is explained by the fact that the trained model can capture the full device dynamics and thus provide optimal pulses that account for spurious effects such as pulse distortions.
As opposed to the physical ME model, the ML model is providing an effectively unbiased estimator of the true quantum dynamics, which for the same level of precision can yield better performing controls in practice.

\section{Conclusion}\label{sec:Conclu}
In this work, we have presented an experimentally simple two-step approach to quantum control aimed at successfully realizing optimal quantum operations in practice.
We tackled the problem of model bias by learning a parametrized representation of the system dynamics directly from experimentally available data, before using that representation to find optimal controls.
Using accurate transmon simulations, we have demonstrated that our machine-learning based quantum optimal control (MLQOC) approach can leverage a practical amount of experimental data to provide a precise characterization of the qubit dynamics and produce high-fidelity ($\sim$~\!99.99\,\%) controls for arbitrary single-qubit gates, thus going significantly beyond realizing a single operation with high fidelity.

The MLQOC framework promises to enable many quantum optimal control applications where model bias is an important bottleneck, without necessitating any real-time feedback control or having to rely on data-intensive blackbox optimizations of single operations.
Beyond experimental ease of implementation, our approach is applicable to a wide range of quantum control problems because its modular features can be adapted to the specificities and requirements of a given quantum system.
Importantly, our method of reducing the complexity of the learning task to only the quantum characterization, together with the use of powerful gradient-based QOC techniques, suggests it might be more data efficient and more scalable to complex control problems than model-free approaches.
It will be interesting to explore how these approaches perform as we expand towards optimally controlling multi-qubit entangling and parallel operations.

\begin{acknowledgments}
    We thank Pierre Guilmin, Ronan Gauthier, and Boris Varbanov for insightful discussions.
    This work is supported by a collaboration between the U.S. Department of Energy and other agencies. This material is based upon work supported by the U.S. Department of Energy, Office of Science, National Quantum Information Science Research Centers, Quantum Systems Accelerator. Additional support is acknowledged from the Natural Sciences and Engineering Research Council, the Canada First Research Excellence Fund, and the Minist\`ere de l’\'Economie et de l’Innovation du Qu\'ebec.
\end{acknowledgments}

\addcontentsline{toc}{section}{References}
\bibliography{biblio}{}

\appendix

\section{Data requirements}\label{appendix:data-requirements}
The success of our approach to quantum optimal control relies on constructing an accurate model of the quantum dynamics from data. We explore here how the amount and quality of training data impact our results.  
We change the accuracy of the data labels by numerically sampling more measurements ($N_\mathrm{shots}$), from the experimentally inexpensive 32 shots (results of the main text), up to the limit of perfect labels. See~\cref{appendix:datasets} for a discussion on different approaches to sampling the pulse shapes we use for training.

\begin{figure}[t]
    \centering
    \includegraphics[width=\columnwidth]{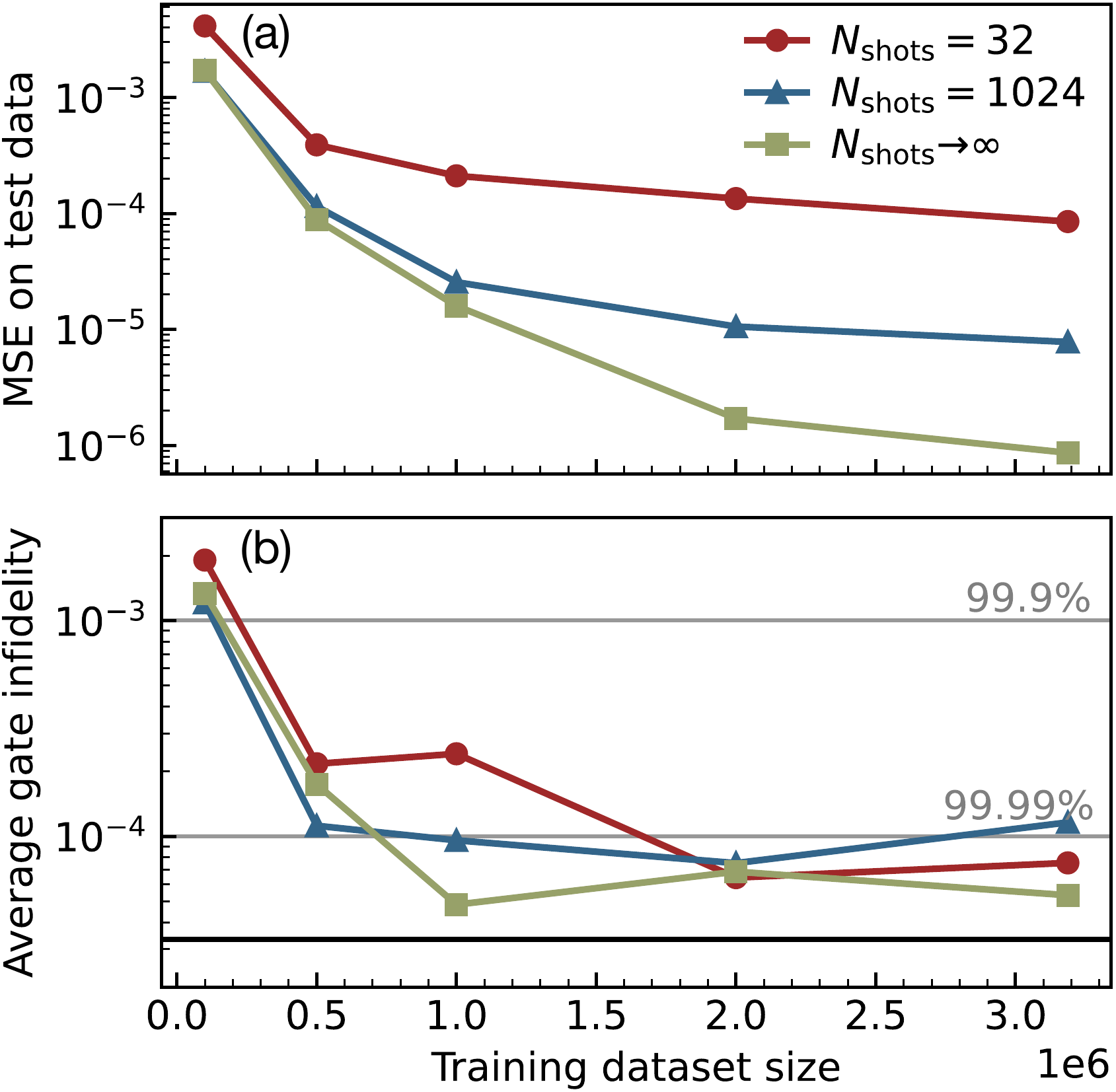}
    \caption{
        Impact of the quantity and quality of training data on the model learning and optimal control performance.
        \textbf{(a)} Mean-squared prediction error on the unseen test dataset of RNN models trained on increasingly more pulse shapes with different sampling repetitions ($N_\mathrm{shots}$).
        \textbf{(b)} Average gate infidelity of \SI{20}{\nano\second} $R_X(\pi)$ gate optimized on the models of panel (a), as evaluated on the true 5-level transmon model. The black line is the coherence limit of the gate.
    }
    \label{fig:App-MLQOC-datasize}
\end{figure}

In~\cref{fig:App-MLQOC-datasize}, we present the impact of the training dataset size and experiment repetitions ($N_\mathrm{shots}$) used on both tasks considered in this work: the ML-based characterization of the quantum dynamics in panel (a) and the resulting quantum optimal control gate fidelity in panel (b). 
Unsurprisingly, using more training data leads to models with better prediction accuracy on the unseen test data.
However, we show that such an improvement in the ML model is not necessarily needed to obtain high-fidelity quantum controls. 
For example, we can get the MSE on the model prediction down by a factor greater than 2 (from $2.1\times10^{-4}$ to $8.5\times10^{-5}$) using 3.2 million training pulse shapes instead of 1 million.
However, the optimal controls found by both of these models lead to a single-qubit gate fidelity of about 99.99\,\%, with most of the variability in the resulting fidelities explained by the stochastic nature of both the model training and the control optimization.
Similarly, increasing the number of shots leads to significant improvements in the model prediction accuracy (about an order of magnitude reduction in the MSE at 3.2 million pulses between the three cases considered), but this improvement does not translate into significant improvements in the resulting control fidelities.

We explain this feature of our approach from the fact that even though the ML model has a finite precision, which can be thought of as error bars on the model parameter estimates, as long as the learned parametrization does not lead to a significant bias in the quantum dynamics, the optimal controls on the finite precision model should also be close to optimal on the true quantum system. We can think of the QOC optimization on the trained neural network, which has a finite precision but very small bias, as adding Gaussian noise to the quantum state evolved using the master equation. The controls that maximize the gate fidelity on average using such a noisy model should also be optimal on the true noiseless model. Here, we show that as long as we have enough data to train a model with a mean-squared prediction error of less than about $2\times10^{-4}$, we can obtain quantum gates with 99.99\,\% fidelity.
Given the important time cost associated with acquiring large amounts of data on the quantum system, it is then desirable to use as little data as required for the MLQOC protocol to yield the desired control fidelities.

We emphasize that on the \SI{}{\mega\hertz}-scale repetition rate typical of quantum systems such as superconducting qubits, only a few hours are required to acquire the amount of data considered here. 
Given that our framework can be easily extended to parallel single-qubit gates with virtually no additional experimental time, optimizing quantum operations using MLQOC is easily achievable with current experiments. 
Additionally, the limited data requirements we obtain for performing arbitrary single-qubit operations open the door to realistically using MLQOC for more complex control tasks such as two-qubit gates, readout, and reset operations, which is beyond the scope of this work.

\section{Machine-learning trainings}\label{appendix:MLtrainings}
In this section, we present the implementation details of our machine-learning (ML) model trainings, whose architecture is illustrated in~\cref{fig:Machine-learning}. Our numerical implementation is based on PyTorch~\cite{pytorch_2019}. 
The recurrent neural network we use is the vanilla long-short term memory unit implemented in \texttt{torch.nn.lstm} with a single layer and a hidden size of 48, see the PyTorch documentation for more details.
The \textit{Encode} transformation is a depth-2 fully connected neural networks (FCNN) with a hidden size of 96 and a sigmoid activation function, whereas the \textit{Decode} transformation is a single-layer linear network, using again a sigmoid activation function to map the network output into proper probabilities $\in(0,1)$.
We emphasize that none of these hyperparameters were carefully optimized, as the objective in this work is to demonstrate how our proposed MLQOC framework can be successful with a simple implementation.
Such an hyperparameter optimization, together with using more powerful neural network architectures such as transformer models~\cite{vaswani_attention_2017}, could yield slightly better performances, but would not change any of our conclusions.

We train this model using gradient-descent based on the Adam optimizer~\cite{kingma_adam_2017} with a learning rate of 0.001. During training, we use an initial batch size of 256 and double it at epochs $[100,200,500,800]$ in order to reduce the stochastic noise in the optimal model parameters as training progresses. 
We use the following physics-inspired loss function to train our model~\cite{genois_quantumtailored_2021}
\begin{align}
    \mathcal{L} &= \mathcal{L}_\mathrm{pred} + w_\mathrm{posit}\mathcal{L}_\mathrm{posit} + w_\mathrm{prep}\mathcal{L}_\mathrm{prep},\\
    \mathcal{L}_\mathrm{pred} &= \frac{1}{N}\sum_{n=1}^N \left((\Pi_n^j-\tilde\Pi_n^j)^2\right),\\
    \mathcal{L}_\mathrm{posit} &= \frac{1}{N(N_t+1)}\sum_{t=0}^{N_t} \sum_{n=1}^N \mathrm{ReLU}\left(\left|\tilde{\rho}_{t,n}\right|^2 - 1 \right),\\
    \mathcal{L}_\mathrm{prep} &= \frac{1}{N}\sum_{n=1}^N \left|\tilde{\rho}_{0,n} - \rho_{0,n}\right|^2,
\end{align}
where $\tilde{\rho}_{t,n}=(\tilde{x}_{t,n}, \tilde{y}_{t,n}, \tilde{z}_{t,n})$ is the model prediction of the qubit states for the $n$th quantum trajectory at time index $t$, given $N$ pulse shapes with $N_t$ pixels. We use the hyperparameters $w_\mathrm{posit}=w_\mathrm{prep}=1.0$.

The prediction loss $\mathcal{L}_\mathrm{pred}$ is a simple mean squared error between the labels and the model predictions for the qubit population in the $j\in\{X,Y,Z\}$ basis, given by $\tilde\Pi_n^j=(1-\widetilde{\langle\sigma^j\rangle})/2\in[0,1]$. Here the labels are also probabilities of measuring the qubit state in the $-1$ eigenstate of the operator $\sigma^j$, which are constructed from averaging the $N_\mathrm{shots}=32$ projective measurements for each input pulse shape.
The MSE given by this loss term is the metric we care most about to obtain an accurate quantum characterization of the qubit dynamics. However, there is more information about the RNN output that we can use to quantify if it accurately describes our physical problem. 

We leverage that information with the following two loss terms, which allow to make our model training more accurate and more data efficient.
The positivity loss $\mathcal{L}_\mathrm{posit}$ insures that the ML model predictions correspond to valid quantum states, i.e. that the predicted density matrices are positive.
The rectified linear unit $\mathrm{ReLU}(x)=\operatorname{max}(0,x)$ allows us to penalize only for nonphysical states and not for mixed state predictions, given that we are learning from data with relaxation and dephasing.
Finally, the preparation loss $\mathcal{L}_\mathrm{prep}$ makes the RNN predictions accurate at the beginning of the quantum trajectory, $t=0$, by using our knowledge of the finite set of prepared initial states.
When dealing with experimental data, or more generally with data containing state preparation and measurement (SPAM) errors, we can set the targeted prepared states $\rho_{0,n}$ to be consistent with our best estimate of these errors. For example, we can average over subsets of the data where the projective readout immediately follows the preparation, and effectively perform quantum state tomography of the six cardinal states of the Bloch sphere~\cite{paris_quantum_2004}.

We note that the model architecture, together with the loss functions we use, can easily be extended to modeling the dynamics of $n$ qubits or multi-level systems. For example, the input dimension of the RNN could be $2n$ for $n$ microwave drives with two quadratures, and the prepared and output quantum states can be represented using $4^n-1$ expectation values for $n$ qubits. This natural extension would be sufficient to apply our framework to optimize for two-qubit gates and parallel single-qubit gates. 
Of course, the data requirements and model complexity necessary for successfully applying MLQOC to controlling a quantum system with an exponentially large Hilbert space should also scale unfavorably with the number of qubits. 
However, using the heuristic representation of a machine-learning model might be beneficial for tackling such complex control problems.

\section{Pulse datasets}\label{appendix:datasets}
The quantum characterization task is achieved by training a parametrized representation, such as a neural network acting as a universal function approximator, to describe the transformation from arbitrary input pulse shapes into the dynamics of the quantum systems under study. Given the enormous size of the input control parameter space, with every pixel taking an arbitrary floating point value, and the fact that we want to acquire a finite dataset in a reasonable experimental time, it is natural to try uniformly sampling at random the input controls to form our ML datasets. This approach works well and the trained ML model can predict the dynamics for random pulse shapes accurately. However, with that choice the model performs poorly at predicting the dynamics of smooth pulses, which for example possess a very different frequency spectrum than random pulses (data not shown). Given that in an experimental scenario we want to use smooth control pulses, such as to respect the bandwidth limitations of the control electronics and avoid crosstalk problems arising from having signals with a wide frequency spectrum, having a trained model that is inaccurate at capturing the dynamics of smooth pulses is practically not useful.

To remedy the situation, we simply include smooth pulses in the training data. Here, we generate these smooth pulses using physically-motivated envelopes that are commonly used to perform single-qubit gates. Specifically, in addition to uniformly sampled random envelopes, our set of pulses is composed of flat, gaussian, gaussian with a flat top, gaussian with an orthogonal DRAG component, and sinusoidal pulses. Heuristically, we chose proportions of 25\,\% of flat pulses with random amplitudes, 25\,\% of flat-top gaussian pulses with random widths, amplitudes, and standard deviations, 25\,\% of uniformly sampled random pulses, 12.5\,\% of gaussian pulses with an orthogonal DRAG component with random amplitudes, and 12.5\,\% of sinusoidal pulses with random amplitudes and frequencies.
We generated a total 4.25 million pulses, and used 75\,\% for the training set, 15\,\% for the validation set used to avoid overfitting, and 10\,\% for the test set used to evaluate and compare model performances.

Given our pulse parametrization at the level of the AWG before IQ mixing, our pulses should have a main sinusoidal oscillation corresponding to the intermediate frequency, here around \SI{100}{\mega\hertz}, such that the resulting pulses reach the qubit frequency and produce non-trivial dynamics. 
We thus generate all of the aforementioned pulses at the envelope level, i.e.~without any explicit oscillating component. We then convolve these envelopes with the intermediate frequency oscillations, which result in the final pulses of the dataset. In order to explicitly sample the impact of detuned drives on the qubit dynamics, we randomly sample the intermediate frequency we use to convolve with our envelopes, using a gaussian distribution centered at the nominal $\omega_\mathrm{IF}/2\pi=$~\SI{100}{\mega\hertz} with a standard deviation of \SI{1}{\mega\hertz}.

We note that fine tuning the quality of the training dataset by using pulses that are known to achieve the target operations with good fidelity, and perhaps increasing the sampling ($N_\mathrm{shots}$) for these pulses, could be beneficial to the performance of the MLQOC approach. However, as we have demonstrated in the main text, such fine tuning is unnecessary and the approach works well as long as the pulses seen during training have the same characteristics as the controls we implement in the end.
In that sense, although such an informed construction of the training dataset will necessarily bias the model learning towards a given set of dynamics, this bias should be beneficial for learning a good model efficiently, as long as the bias is consistent with the constraint we put on the following QOC optimizations. Indeed, it is desirable to have a trained model that is good at predicting the dynamics of smooth pulses, even if it is somewhat specialized and not fully generalizable, since ultimately this is what our optimal control task requires. This bias is enforced explicitly in this work by using a cost function minimizing the first and second derivatives of the optimal controls.

\section{QOC implementation details}\label{appendix:QOC}
In this section, we present the implementation details of our quantum optimal control (QOC) approach illustrated in~\cref{fig:Overview}(b).
We use an open-loop optimization where the pulses are parametrized at every pixel, the trained ML model with fixed parameters is used to compute the dynamics resulting from the pulses, and the cost function is expressed as follows~\cite{leung_speedup_2017}
\begin{align}
    \mathcal{C} &= \mathcal{C}_\mathrm{fidel} + w_\mathrm{clamp}\mathcal{C}_\mathrm{clamp} + w_\mathrm{mean}\mathcal{C}_\mathrm{mean}\\
    &+ w_\mathrm{first}\mathcal{C}_\mathrm{first} + w_\mathrm{second}\mathcal{C}_\mathrm{second},\\
    \mathcal{C}_\mathrm{fidel} &= \frac{1}{N}\sum_{n=1}^N 1-\operatorname{AGF}\left(\rho_{n,T}\right),\\
    \mathcal{C}_\mathrm{clamp} &= \frac{1}{N(N_t+1)}\sum_{t=0}^{N_t} \sum_{n=1}^N \mathrm{ReLU}\left(\Omega_{n,t} - \Omega_\mathrm{max} \right),\\
    \mathcal{C}_\mathrm{mean} &= \frac{1}{N(N_t+1)}\sum_{t=0}^{N_t} \sum_{n=1}^N \left|\Omega_{n,t}\right|^2,\\
    \mathcal{C}_\mathrm{first} &= \frac{1}{N N_t}\sum_{t=1}^{N_t} \sum_{n=1}^N \left|\partial_t\Omega_{n,t}\right|^2,\\
    \mathcal{C}_\mathrm{second} &= \frac{1}{N(N_t-1)}\sum_{t=2}^{N_t} \sum_{n=1}^N \left|\partial_t^2 \Omega_{n,t}\right|^2,
\end{align}
where $\Omega_\mathrm{max}$ is the maximal allowed drive amplitude, which is $2\pi\times$\SI{100}{\mega\hertz} in this work. This constitutes a reasonable choice to avoid significantly amplifying the effect of classical crosstalk on a transmon chip, as typical sinusoidal \SI{20}{\nano\second} single-qubit $\pi$ gates require $|\Omega|/2\pi\approx$~\SI{31}{\mega\hertz}.
The partial time derivatives in the smoothing cost functions are implemented as finite differences numerically with $\partial_t\Omega_{n,t} = \Omega_{n,t} - \Omega_{n,t-1}$, and the average gate fidelity is computed as~\cite{nielsen_simple_2002}
\begin{align}
    \operatorname{AGF}\left(\rho_{n,T}\right) = \frac{\sum_j\operatorname{Tr}\left(U_\mathrm{target} \rho^j U_\mathrm{target}^\dagger \operatorname{RNN}(\rho^j)\right) + d^2}{d^2(d+1)},
\end{align}
where $d=2$ for a qubit, $T$ the final time index, $\sigma^j$ are the 6 cardinal states (our chosen unitary 2-design), and $\operatorname{RNN}(\rho^j)$ represents the transformation realized by our trained ML model, which acts here in place of the usual quantum channel.

As mentioned in the main text, we optimize for a batch of 30 pulses in parallel on the same GPU card in under a minute, and select the best performing pulse on the full cost function as the optimal control. We use the Adam optimizer with a learning rate of 0.001 to perform the gradient descent~\cite{kingma_adam_2017}. As in the ML model training, all the gradients are computed numerically using the auto-differentiable feature of PyTorch~\cite{pytorch_2019}. 

We have observed that whereas the clamping cost $\mathcal{C}_\mathrm{clamp}$ simply allows the pulses to respect an experimental constraint, the average amplitude cost ${C}_\mathrm{mean}$ together with the smoothing cost functions $\mathcal{C}_\mathrm{first}$ and $\mathcal{C}_\mathrm{second}$ are very important for constraining the parameter exploration during the optimization towards controls where the trained RNN has an accurate representation of the dynamics.
Indeed, the trained model predicts dynamics which are very close to the true master equation dynamics for smooth input pulses, which allows us to find optimal controls with over 99.99\,\% fidelity when using these regularizing cost functions.
However, optimizing the controls on the trained model using only the fidelity cost can lead to pulses where the predicted and true dynamics diverge significantly. This problem can be completely avoided using the smoothing cost functions, and if necessary a simple post-selection based on the same smoothness criteria over the optimization results of the batch.
We optimize our pulses for all the single-qubit gate results presented in the main text using $w_\mathrm{clamp}=10.0$, $w_\mathrm{mean}=0.1$, and $w_\mathrm{first}=w_\mathrm{second}=0.01$. We have observed that fine tuning these hyperparameters do not lead to significant changes in the resulting optimal gate fidelities.

\section{Random pulse distortions}\label{appendix:distortions}
We simulate pulse distortions numerically by applying a fixed causal transfer function to the pulses. We first generate the random transfer function $F(\omega)$ in the frequency domain, using the noise parameter $\lambda$ that we scale from 0 to 0.4 in~\cref{fig:Robustness}.
We add the noise to the trivial flat transfer function such that
\begin{align}
    F(\omega\geq0) = 1 + U(-\lambda, \lambda),
\end{align}
where $U(\mathrm{min}, \mathrm{max})$ is the uniform distribution, and we replicate the values for the negative frequencies such that $F(\omega)$ is even.  
To obtain a smooth transfer function, we sample 11 points from the uniform distribution, that we attribute to frequencies up to \SI{600}{\mega\hertz}, before using a gaussian filter with a \SI{2.5}{\giga\hertz} bandwidth to interpolate between these values for 1001 points in the range $[-1.5,1.5]$~\SI{}{\giga\hertz}.

Instead of using Fourier transforms to convolve all the pulses of our datasets with the noisy transfer function defined in the frequency domain, we express $F(\omega)$ in the time domain using a transfer matrix which can be directly applied to the pulses using a simple matrix multiplication. This transfer matrix is obtained by numerically solving~\cite{motzoi_optimal_2011}
\begin{align}
    T_{jk} = \int_{-\infty}^\infty F(\omega) \frac{\sin(\omega\Delta t /2)\cos(\omega(j-k)\Delta t)}{\pi\omega}\mathrm{d}\omega,
\end{align}
where $\Delta t$ is the pixel size (in units of time) and $j,k$ are the pixel indices. 

Finally, given that this noise models a physical process, we impose causality by setting the lower triangular elements to zero and renormalizing it. Using this approach, we can simulate a realistic scenario where out-of-model dynamics are present in the quantum system we are trying to control, and we can have a single tunable parameter $\lambda$ to quantify the resulting model bias.

\section{IQ mixing and qubit drive}\label{appendix:IQmixer}
In this section, we detail the IQ mixing transformation applied to the input microwave pulses, typically generated by an arbitrary waveform generator (AWG), before reaching the transmon qubit. We then show how these mixed pulses can be viewed as the usual $\sigma_X$ and $\sigma_Y$ drive Hamiltonian prefactors.
As detailed in Refs.~\cite{krantz_quantum_2019,baur_realizing_2012}, the two AWG signals $S_I$ and $S_Q$ get mixed with a local oscillator signal of frequency $\omega_\mathrm{LO}$ in a process known as sideband mixing to produce the drive
\begin{align}
    \Omega(t) = S_I(t) \cos(\omega_\mathrm{LO}t) + S_Q(t) \sin(\omega_\mathrm{LO}t).
\end{align}
We then define the AWG signals with an envelope combined with sinusoidal oscillations at an intermediate frequency $\omega_\mathrm{IF}$ and a phase $\phi$, such that 
\begin{align}
    S_I(t) &= I(t) \cos(\omega_\mathrm{IF}t+\phi),\\
    S_Q(t) &= -Q(t) \sin(\omega_\mathrm{IF}t+\phi).
\end{align}
We then directly get
\begin{align}
    \Omega(t) &= \frac{I(t)}{2} \left[\cos(\omega_+t + \phi) + \cos(\omega_-t + \phi)\right]\\
    &- \frac{Q(t)}{2} \left[\cos(\omega_-t + \phi) - \cos(\omega_+t + \phi)\right]
\end{align}
where $\omega_\pm=\omega_\mathrm{IF}\pm\omega_\mathrm{LO}$.
We thus see that by setting $I(t)=Q(t)=A(t)$, we obtain a single carrier frequency signal at the upper sideband $\omega_+$,
\begin{align}
    \Omega(t) &= A \cos(\omega_+t +\phi).
\end{align}
Using a LO close to the qubit frequency such that $\omega_\mathrm{IF}+\omega_\mathrm{LO}\approx\omega_q$, we can then precisely control the frequency spectrum of the qubit pulses we send, with a limitation set by the AWG bandwidth and potentially the cable filters used in the experiment. 
Note that one can also choose a different phase between the $S_I$ and $S_Q$ signals to drive the lower sideband.

We now demonstrate that such a drive can be used to perform arbitrary qubit gates. 
Starting from~\cref{eq:hamiltonian}, we can introduce the creation and annihilation operators to diagonalize the quadratic terms of the transmon Hamiltonian and obtain, after expanding the $\cos\hat\phi$ term and truncating after the second order~\cite{blais_RMP_2020}
\begin{align}
    \hat H(t) \approx \omega_q\bdag\bb - \frac{E_C}{2}\bdag\bdag\bb\bb + \Omega(t) \frac{i}{2} \left(\frac{2E_C}{E_J}\right)^{1/4} (\bdag - \bb).
\end{align}
Going into the qubit rotating frame with the transformation $\hat U(t) = \exp(i \omega_q \bdag\bb t)$ and applying the rotating-wave approximation (RWA), we obtain the time-dependent drive Hamiltonian
\begin{align}
    \hat H_d(t) \approx \frac{i}{2} \left(\frac{2E_C}{E_J}\right)^{1/4} \frac{A(t)}{2} \left(\bdag e^{i(\Delta t+\phi)} - \bb e^{-i(\Delta t+\phi)}\right),
\end{align}
where $\Delta=\omega_+ - \omega_q$.
Performing a two-level truncation with $\bdag\to\sigma_+$, $\bb\to\sigma_-$, and $\sigma_\pm=(\sigma_X\pm\sigma_Y)/2$, redifining the prefactor in front of the last parenthesis to be $\tilde A(t)$ and assuming that the drive is on resonance with the qubit ($\Delta=0$), we finally get the textbook single-qubit drive Hamiltonian
\begin{align}
    \hat H_{d}^\mathrm{qubit}(t) = \tilde{A}(t) \left[\cos(\phi)\sigma_X - \sin(\phi)\sigma_Y \right].
\end{align}
We thus understand how controlling the pixel amplitudes of the AWG inputs $S_I(t)$ and $S_Q(t)$ allows us to perform arbitrary single-qubit operations.

\end{document}